\documentclass[aps,pre,amsmath,amssymb,superscriptaddress,reprint]{revtex4-2}
\pdfoutput=1
\usepackage[T1]{fontenc}
\usepackage{graphicx}
\makeatletter

\usepackage[english]{babel}
\usepackage[T1]{fontenc}
\usepackage[utf8]{inputenc}
\IfFileExists{lmodern.sty}{\usepackage{lmodern}}{}
\usepackage{color}
\definecolor{darkblue}{rgb}{0,0,0.6}
\definecolor{darkred}{rgb}{0.6,0,0}
\definecolor{darkgreen}{rgb}{0,0.6,0}
\usepackage{setspace}
\usepackage{environ}
\usepackage{wasysym}
\usepackage{times}
\usepackage[colorlinks=true,urlcolor=darkblue,citecolor=darkblue,linkcolor=darkred,hyperindex=true,hyperfootnotes=false]{hyperref}
\usepackage{mathtools}
\usepackage{mathrsfs}
\usepackage{bm}
\newcommand{\setword}[2]{%
  \phantomsection
  #1\def\@currentlabel{\unexpanded{#1}}\label{#2}%
}

\newcommand{\Deff}{D_{\rm eff}}
\newcommand{\bj}{\mathbf{J}}
\newcommand{\br}{\mathbf{r}}
\newcommand{\by}{\mathbf{y}}
\newcommand{\bnabla}{\bm{\nabla}}
\makeatletter

\begin{document}
\title{Universal correlations in local measurements directly probe effective diffusivity}

\author{\vspace*{-2mm}Omer Granek}
\affiliation{Leinweber Institute for Theoretical Physics \& Kadanoff Center for Theoretical Physics, University of Chicago, 933 E 56th St, Chicago,  Illinois 60637, USA}
\altaffiliation{Present address.}
\affiliation{Department of Physics,
Technion -- Israel Institute of Technology,
Haifa, 3200003, Israel}

\begin{abstract}
Measuring transport coefficients at the microscale remains challenging, often relying on indirect methods that require modeling and calibration. This article derives universal asymptotic forms for the autocorrelation and relative uncertainty of local probe measurements in dilute diffusive systems. Valid both at and far from equilibrium, these forms directly connect microscopic measurements to the effective diffusivity. Indirect methods such as dynamic light scattering and fluorescence correlation spectroscopy can therefore serve as asymptotically direct probes applicable to active and other microscopically nondiffusive systems. Simulations across several models and observables confirm the broad applicability of these predictions to a variety of probes.
\end{abstract}

\maketitle
\section{Introduction}
Local observables such as temperature and current can be measured directly using probes such as thermometers and ammeters.  Modern technological advances have pushed such probes to the micro and nano scales: colloids trapped by optical tweezers can act as microscopic thermometers~\cite{Pesce2020,Romero-Gonzalez2023}, while natural and artificial ion channels serve as nanoscopic ammeters~\cite{Crescentini2014,Yan2025}. Miniaturization of local probes has therefore enabled access to ever smaller length scales.

Effective transport coefficients, by contrast, are inherently macroscopic. Matter transport is governed by the number conservation law~\cite{Mehrer2007},
\begin{align}
    \partial_t \rho(\br,t) = -\bnabla\cdot\bj(\br,t)\;,\label{eq:conserv}
\end{align}
where $\rho(\br,t)$ and $\bj(\br,t)$ are the average number and current density fields, respectively. Diffusive systems form a broad class of microscopic dynamics that, on large scales, converge to Fick's law of diffusion $\bj\!\sim\!-\Deff\bnabla\rho$, which defines the effective diffusivity $\Deff$~\cite{Tyrrell1984,Kubo1991,Derrida2007,Mehrer2007}. In the dilute limit where interactions can be neglected, $\Deff$ is equivalently defined through the mean squared displacement (MSD) of a tagged particle, which follows Einstein's law $\langle \mathbf{r}^2(t)\rangle\!\sim\! 2d \Deff t$ at long times $t$, with $d$ the dimension and $\langle \cdot \rangle$ an average over histories~\cite{Tyrrell1984,Kubo1991,Mehrer2007}. Diffusion is ubiquitous both at and far from equilibrium, from electrons in semiconductors~\cite{VanDerZiel1978,Bonani2001}, atoms in solids~\cite{Mehrer2007} and colloids in fluids~\cite{dhont1996introduction}, to transport in porous media~\cite{Webb2003,Tartakovsky2019}, heterogeneous biomaterials~\cite{Loren2009}, and active baths~\cite{Jee2018,Feng2020,Ghosh2021,Granek2024}. In all these settings, direct access to $\Deff$ is limited to sufficiently large scales where coarse-grained diffusion holds~\cite{Tyrrell1984,Mehrer2007}: Fick's law can be measured directly from the global response to inhomogeneous boundary or initial conditions~\cite{Gordon1945,tanford1961physical,Rutherford1997,Lee2017,Nguyen2022,Hamada2023}. Likewise, Einstein's law can be measured directly from extended particle trajectories using single-particle tracking~\cite{Manzo2015,Shen2017,Kumar2023} or NMR spectroscopy~\cite{Price2009,Karger2013}.

Nevertheless, $\Deff$ can be measured microscopically using indirect methods such as dynamic light scattering (DLS)~\cite{Berne2000,Stetefeld2016}, fluorescence correlation spectroscopy (FCS)~\cite{Krichevsky2002,Lakowicz2006,Gunther2018,Yu2021}, and fluorescence recovery after photobleaching~\cite{Loren2015,Pincet2016}, which are now widely used in science and industry. DLS and FCS estimate $\Deff$ by fitting the normalized autocorrelation function (ACF),
\begin{align}
    G(t)\equiv \frac{\langle A(t)A(0)\rangle_c}{\langle A\rangle^2}=\frac{\langle A(t)A(0)\rangle-\langle A\rangle^2}{\langle A\rangle^2}\;,\label{eq:Gdef}
\end{align}
where $A(t)$ is the recorded intensity of light scattered off the particles (DLS) or emitted by bound fluorophores (FCS) in a localized detection region.
Extracting $\Deff$ from $G(t)$ requires modeling and calibration, often relying on the assumption that Fick’s law already holds at the probe scale~\cite{Berne2000,Lakowicz2006}. This assumption becomes fragile in microscopically nondiffusive systems, including active baths~\cite{Boon1974,Wilson2011,Gunther2018}, where the probe is precisely most sensitive to internal dynamics. Despite a broad interest in the effective diffusivity of active baths~\cite{Jee2018,Kandula2019,Jee2019,Feng2020,Jee2020,Ghosh2021,Granek2024}, accessing $\Deff$ without relying on probe-scale diffusion has remained largely unexplored.

This article addresses this gap for dilute, diffusive systems of noninteracting particles in a steady state. A system of size $L$ containing $N$ particles interacts with a fixed probe in the thermodynamic limit $N,L\rightarrow\infty$ at fixed density $\rho_0=N/L^d$. The probe performs an arbitrary local measurement and produces a stationary fluctuating signal $A(t)$  with $\langle A \rangle \neq 0$. The first main result of this article is a universal long-time form for the ACF:
\begin{align}
    G(t)= \frac{1}{\rho_0 (4\pi\Deff t)^{d/2}}+\mathcal{O}(t^{-(d/2+1)})\;.\label{eq:G}
\end{align}
Equation~\eqref{eq:G} follows from Eq.~\eqref{eq:prop} below, derived via the spectral expansion leading to Eq.~\eqref{eq:propspec}. For $d\leq2$, Eq.~\eqref{eq:G} implies a second universal asymptotic form for the squared relative uncertainty $\varepsilon^2(t)\equiv \langle\bar{A}(t)^2\rangle_c/\langle\bar{A}\rangle^2$ of the time-average $\bar{A}(t)\equiv \int_0^t dt'A(t')/t$,
\begin{align}
\!\!\!\varepsilon^{2}(t)\sim\begin{dcases}
\rho_0^{-1}\!\left(1\!-\!\tfrac{3}{4}d\!+\!\tfrac{1}{8}d^2\right)^{-1}\!(4\pi \Deff t)^{-d/2}\,,\!\!&\!d<2\\
\rho_0^{-1}(2\pi \Deff t)^{-1}\log t\,,\!&\!d=2
\end{dcases},\label{eq:eab}
\end{align}
which is the second main result of this article. Equations~(\ref{eq:G}-\ref{eq:eab}) directly relate local measurement fluctuations to the macroscopic properties $\Deff$, $\rho_0$ and $d$. Probe details, including the measured observable, microscopic dynamics, and system-probe interaction, enter only through the subleading $\mathcal{O}[t^{-(d/2+1)}]$ corrections to Eqs.~(\ref{eq:G}-\ref{eq:eab}) (see Appendix). Equations~(\ref{eq:G}-\ref{eq:eab}) hold both at and far from equilibrium for dilute, noninteracting systems with diffusive coarse-grained dynamics.
While the exponents in Eqs.~(\ref{eq:G}-\ref{eq:eab}) are expected in diffusive systems~\cite{Erdos1951,Redner2001,Majumdar2024}, the nontrivial universality lies in the prefactors. The latter yield a local definition of $\Deff$ in the dilute regime, commonly assumed in DLS and FCS analyses~\cite{Berne2000,Lakowicz2006}, and enables \emph{direct} measurement of $\Deff$ in this limit.

\section{Connection to standard DLS and FCS formulae}
As a concrete example, Eq.~\eqref{eq:G} is shown to agree with the standard theory of FCS, which derives $G(t)$ for noninteracting Brownian particles (BPs) illuminated by a Gaussian beam in $d=3$. The effective detection region is axially symmetric and has a characteristic radius $w$ and aspect ratio $\kappa$. Fluctuations in the detected particle count result in~\cite{Lakowicz2006}
\begin{align}
    \!\!\!G(t) = \langle n\rangle^{-1}\left[1+\frac{t}{\tau(\Deff)}\right]^{-1}\left[1+\frac{1}{\kappa^2}\frac{t}{\tau(\Deff)}\right]^{-1/2},\label{eq:FCS}
\end{align}
where $\tau(\Deff)\equiv w^2/4\Deff$ and $\langle n\rangle=\rho_0 v_{\rm eff}$ is the average particle count in the effective detection volume $v_{\rm eff}=\pi^{3/2}\kappa w^3$. Rescaling time by $\tau(\Deff)$ eliminates $w$, leaving $\kappa$ as a dimensionless fitting parameter~\cite{Hofling2011,Hofling2013,Banks2016}. Nonetheless, extracting $\Deff$ from the measurement requires calibrating $w$ in $\tau(\Deff)$ using a known diffusivity~\cite{Lakowicz2006}. In contrast, the expansion of Eq.~\eqref{eq:FCS} for $t\gg\tau(\Deff)$, together with $\langle n\rangle=\rho_0 v_{\rm eff}$, readily reduces to Eq.~\eqref{eq:G}, eliminating both $w$ and $\kappa$ at leading order.

The conclusion extends to DLS. There, Eq.~\eqref{eq:FCS} is supplemented by the intermediate scattering term $|\mathcal{F}(\mathbf{q},t)|^2\!=\!\beta e^{-2\Deff q^2 t}$, where $\bf{q}$ is the scattering wavevector and $\beta=\mathcal{O}(1)$ is the detector's coherence factor~\cite{Berne2000,Stetefeld2016}. Although $|\mathcal{F}(\mathbf{q},t)|^2$ dominates the initial decay in the Gaussian limit $\langle n\rangle\gg1$, the asymptotic long-time behavior again crosses over to Eq.~\eqref{eq:G}.
More generally, extensions of FCS provide various detection profiles in $d=1,2,3$~\cite{Blom2009}. As demonstrated by Eq.~\eqref{eq:denscorr} below, the experimentally-validated theory of DLS and FCS converges to Eq.~\eqref{eq:G} in the long-time limit for arbitrary probe geometries.

\section{Beyond microscopic diffusion: simulations of active and passive baths}
\begin{figure}
    \centering
    \includegraphics[width=\linewidth]{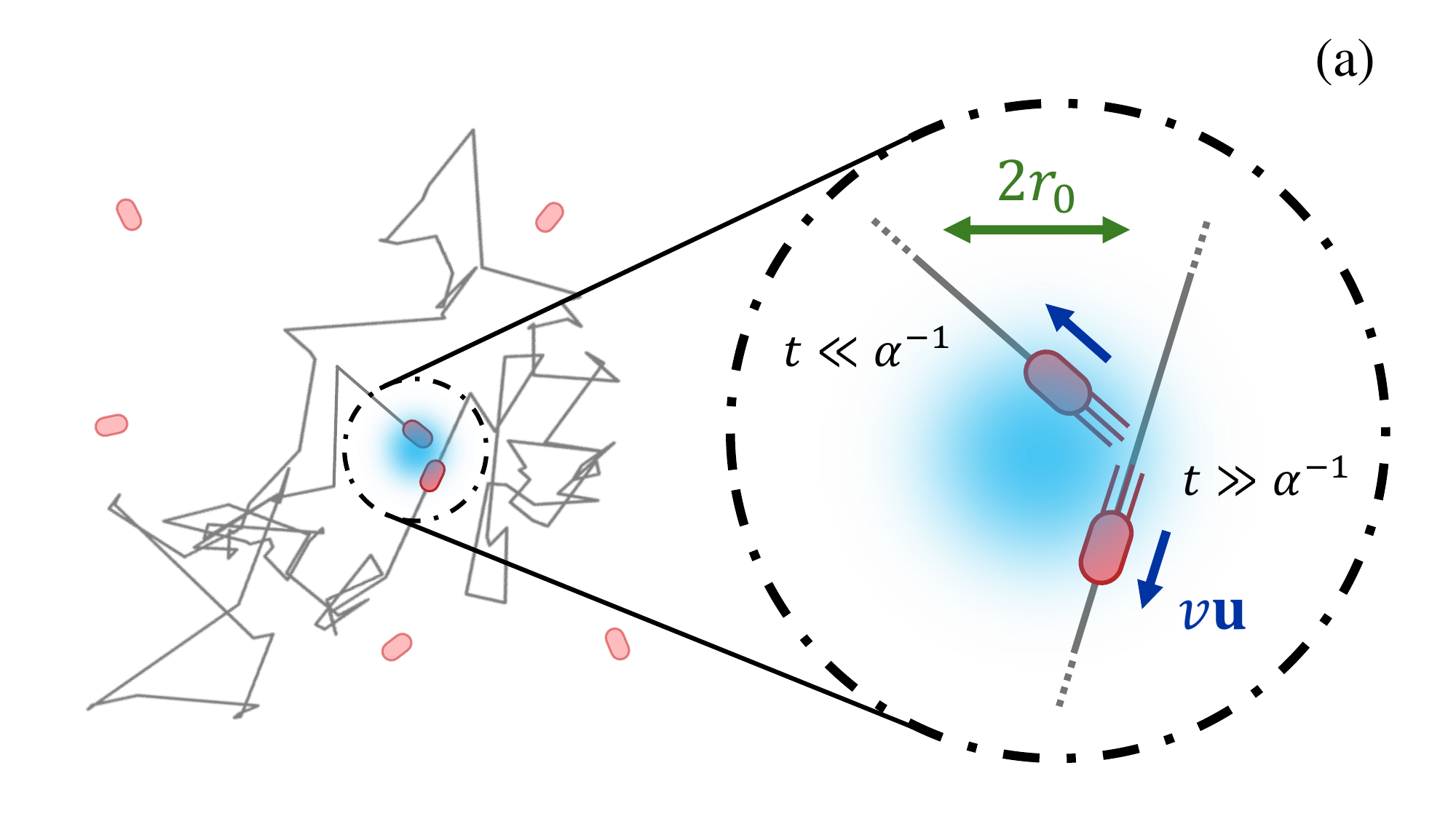}
    \includegraphics[width=\linewidth]{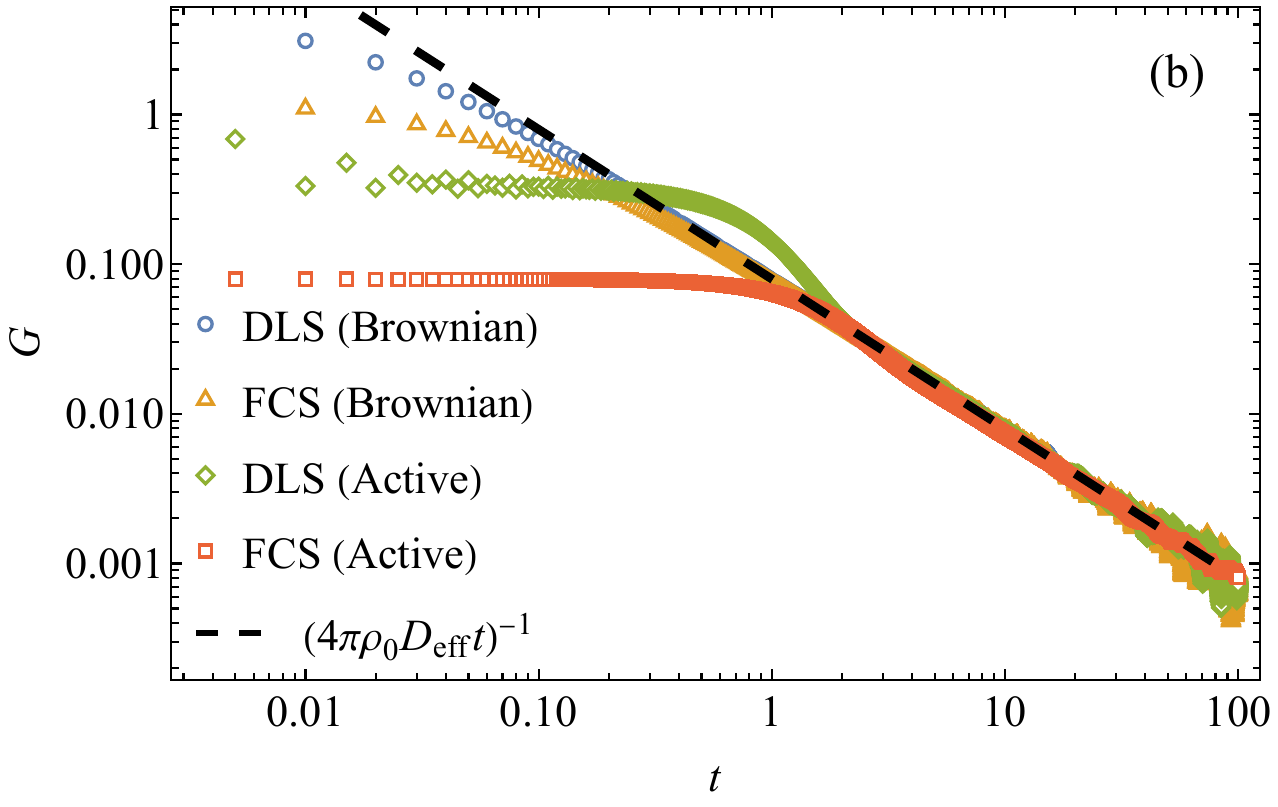}
    \vspace*{-7mm}
    \caption{\textbf{(a)} Sketch of $d=2$ DLS/FCS simulations of RTPs traveling at velocity $v\mathbf{u}$ (blue arrow) and tumbling at rate $\alpha$. The Gaussian detection profile (cyan gradient) has radius $r_0=v/\sqrt{2}\alpha$ (DLS) and $r_0=v/\alpha$ (FCS). The trajectory (gray solid line) is ballistic on the detector scale and diffusive on larger scales.  \textbf{(b)} 
ACF in DLS/FCS simulations of BPs and RTPs in $d=2$ (symbols) and theory, Eq.~\eqref{eq:G} (black dashed line; no fitting parameters). Parameters are set to unity, except: $L=10^2$, $\alpha=1/2$, $\mathbf{q}=(-263,32)$, $E_0=10^6$. ACF curves are operationally measured using time averages over the acquisition time $\mathcal{T}=10^6$~\cite{Berne2000,Lakowicz2006} and further averaged over $8$ (BP DLS) or $6$ acquisitions (rest)~\cite{LogAvg}.}
    \label{fig:DLSFCS}
    \vspace*{-5mm}
\end{figure}

\begin{figure}
\includegraphics[width=1\columnwidth]{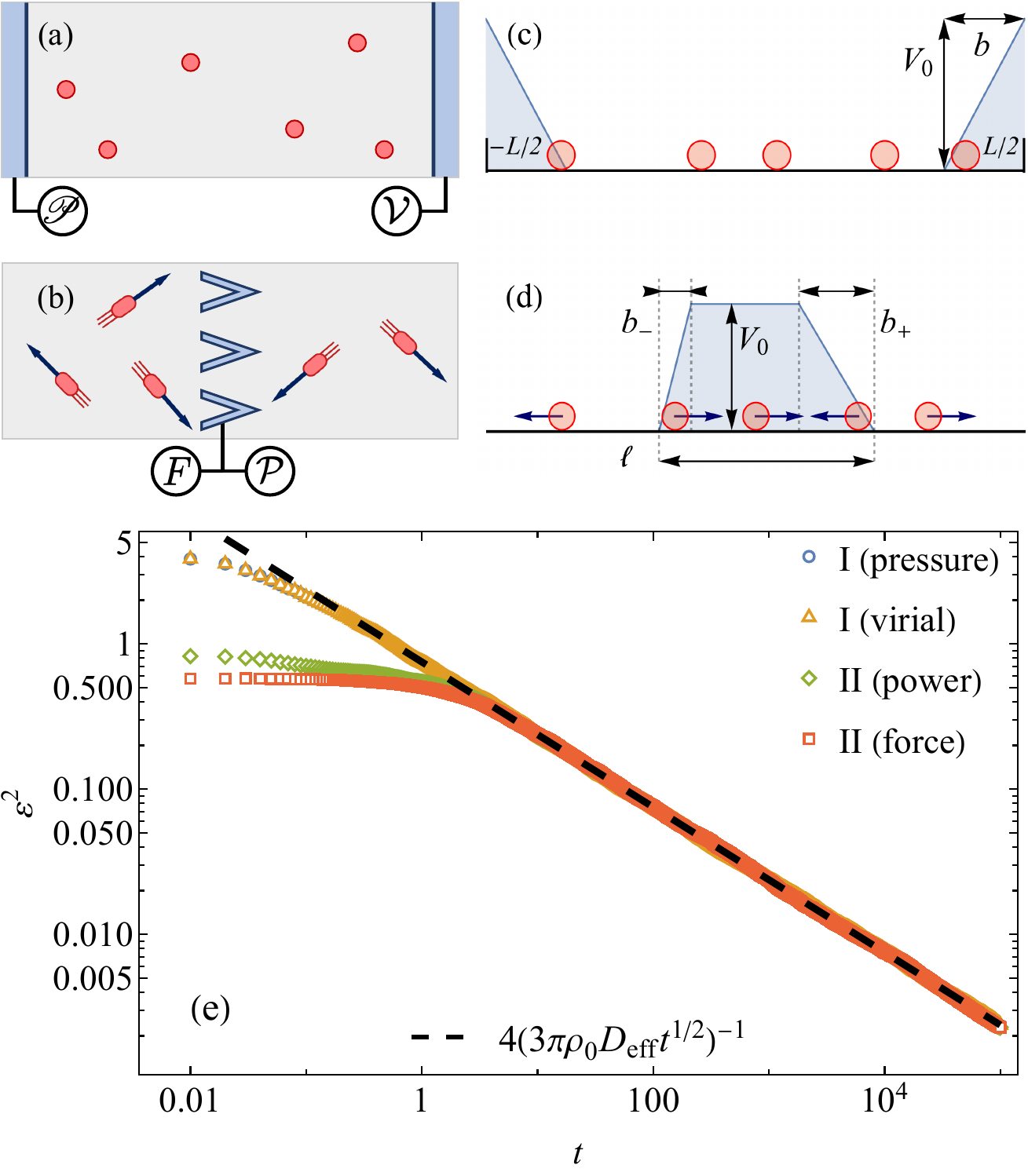}
\vspace*{-7mm}
\caption{\textbf{(a-b)} Sketches of probes in dilute systems. The measurement integrates over a $d=1$ manifold, projecting the $d=2$ system onto an effective $d=1$ space. \textbf{(a)} BPs confined by walls connected to mechanical pressure ($\mathscr{P}$) and virial ($\mathcal{V}$) sensors. \textbf{(b)} Active RTPs interacting with an asymmetric permeable partition connected in its entirety to force ($F$) and power ($\mathcal{P}$) sensors. System (a) is at equilibrium; (b) is far from equilibrium. \textbf{(c-d)} Sketches of the effective $d=1$ Models~\ref{word:I}-\ref{word:II} (respectively).
\textbf{(e)} Squared relative uncertainty of steady-state time-averaged measurements; simulations (colored symbols) and theory, Eq.~\eqref{eq:eab} (dashed black line; no fitting parameters). Pressure and virial symbols overlap. Parameters are set to unity, except: $L=10^4$ and acquisition time $\mathcal{T}=10^5$.  \ref{word:I}: $V_0=30$, $b=3$. \ref{word:II}:
$V_0\!=\!0.25$, $b_-\!=\!0.26$, $b_+\!=\!0.74$.\label{fig:eps}}
\vspace*{-5mm}
\end{figure}

Beyond reproducing the microscopically-diffusive limit, simulations in Fig.~\ref{fig:DLSFCS} show that Eq.~\eqref{eq:G} remains valid even when probe-scale dynamics is strongly nondiffusive and Fick’s law fails locally. Simulations in Fig.~\ref{fig:eps} further show that Eqs.~(\ref{eq:G}-\ref{eq:eab}) apply to other observables beyond standard light signals. The simulation setup is described next. A systematic derivation of Eqs.~(\ref{eq:G}-\ref{eq:eab}) follows, and implications and extensions are finally discussed.

Particles are evolved according to overdamped dynamics in $d=1,2$~\cite{Granek2024},
\begin{align}
    \dot{\mathbf{r}}_n = v\mathbf{u}_n(t)-\mu\bm{\nabla}V[\mathbf{r}_n(t)] +\sqrt{2\mu T}\bm{\eta}_n(t)\;.\label{eq:sim}
\end{align}
Here, $\{\mathbf{r}_n(t)\}_{n=1}^N$ are the particle positions, $\{\mathbf{u}_n(t)\}_{n=1}^N$ are unit vectors, randomized at Poissonian rate $\alpha$, and $\{\bm{\eta}_n(t)\}_{n=1}^N$ are unit-variance white noises. The potential $V(\br)$ describes the interaction with the probe, with $\mu$ being the mobility and $T$ being the ambient temperature. 
The case $v=0$ corresponds to equilibrium BPs with diffusivity given by the Einstein relation $\Deff=\mu T$. The case $T=0$ corresponds to active run-and-tumble particles (RTPs)~\cite{Cates2012}. The RTP models a self-propelled bacterium that moves at constant speed $v$ while randomly reorienting its direction $\mathbf{u}$ through instantaneous ``tumbling'' events at rate $\alpha$~\cite{Schnitzer1993,Cates2012}.

Figure~\ref{fig:DLSFCS} extends FCS and DLS to active baths by comparing simulations of active RTPs and passive BPs in $d=2$ with $V=0$. This minimal setup models the planar measurement of, e.g., active bacteria near surfaces~\cite{Lemelle2020} and passive membrane inclusions~\cite{Lakowicz2006}. The measured signals are $I(t)=\big|\sum_n e_n(t)e^{i\mathbf q\cdot\mathbf r_n(t)}\big|^2$ (DLS) and $I(t)=\sum_n e_n^2(t)$ (FCS), with the detected field $e_n(t)\!=\!E_0\exp\{-[r_n(t)/r_0]^2\}$. In the active case (see Fig.~\ref{fig:DLSFCS}(a)), the RTP dynamics dictate an average persistence length $\ell_{\rm p}=v/\alpha$. On timescales $t\ll\alpha^{-1}$ and length scales $\ell\ll\ell_{\rm p}$, the motion is ballistic, with particles moving in straight lines. For $t\gg\alpha^{-1}$ and $\ell\gg \ell_{\rm p}$, repeated tumbling leads to effective diffusion with $\Deff=v^2/\alpha d$. In the simulation, the detection radius is $r_0=\ell_{\rm p}/\sqrt{2}$ (DLS) and $r_0=\ell_{\rm p}$ (FCS), so Fick’s law does not hold on the measurement scale. Nonetheless, the ACF in all settings converges to Eq.~\eqref{eq:G} in the long-time limit.

Beyond FCS/DLS, Fig.~\ref{fig:eps} demonstrates Eq.~\eqref{eq:eab} using various local observables.
Although Eq.~\eqref{eq:G} holds in any dimension $d$, Eq.~\eqref{eq:eab} applies only to $d\leq2$. For $d>2$, the well-known nonuniversal form $\varepsilon^2(t)\sim2\mathcal{D}/t$ is recovered, where the dispersion $\mathcal{D}$ is given by the Green–Kubo relation $\mathcal{D} \!=\!\int_{0}^{\infty}dt\,G(t)$~\cite{Tyrrell1984,Kubo1991}. Nevertheless, dilute $d>1$ systems effectively become non-interacting $d=1$ systems if the measurement projects the dynamics onto a $d=1$ space. This is done by, e.g., probes connected to walls and partitions, where the measurement signal depends solely on motion normal to the wall. Figure~\ref{fig:eps} thus demonstrates Eq.~\eqref{eq:eab} for two $d=1$ probe models (Fig.~\ref{fig:eps}(a–d)): 
(\textbf{\setword{I}{word:I}}) BPs in a soft box and (\textbf{\setword{II}{word:II}}) active RTPs near an asymmetric permeable partition~\cite{Galajda2007,Tailleur2009a,Nikola2016b}. To further support the universality, the SM includes two additional models~\cite{SM}:  (\textbf{\setword{III}{word:III}}) a localized variant of the Ajdari–Prost flashing Brownian ratchet model~\cite{Ajdari1992,Prost1994,Julicher1997} and (\textbf{\setword{IV}{word:IV}}) random walkers on a lattice, driven by a pointlike pump~\cite{Sadhu2011a}. 

Model~\ref{word:I} is an exact projection of the system in Fig.~\ref{fig:eps}(a) onto the (horizontal) measurement axis~\cite{footnote1}. In this model, $V(x)$ describes the interaction with soft walls of width $b$ and potential height $V_0$ (see Fig.~\ref{fig:eps}(c)). The walls are effectively impenetrable since $V_0/T$ is sufficiently large that the typical crossing time far exceeds the acquisition time $\mathcal{T}$~\cite{pent,SM}. Model~\ref{word:II} is a simplified model for the projection of the system in Fig.~\ref{fig:eps}(b) onto the measurement axis. In this model, $V(x)$ describes the interaction with a penetrable asymmetric probe of total length $\ell$, potential height $V_0$ and side widths $b_\pm$ (see Fig.~\ref{fig:eps}(d))~\cite{SM}.

Two observables are measured for each model. For the equilibrium Model~\ref{word:I}, these are the net pressure $\mathscr{P}(t)\!=-\!\sum_n\!{\rm sgn} (x_n(t))f_n(t)/2$ exerted on the box walls and the Clausius virial $\mathcal{V}(t)\!=-\!\sum_nx_n(t)f_n(t)/2$, where $f_n(t)=-V'[x_n(t)]$ are the forces exerted on each particle. Both are thermodynamic observables that admit the ideal-gas equations of state $\langle\mathscr{P}\rangle\!=\!\rho_0 T$ and $\langle\mathcal{V}\rangle\!=-\!\rho_0 
 L T/2$. For the far-from-equilibrium Model~\ref{word:II}, the observables are the net force $F(t)\!=\!\sum_n f_n (t)$ the probe exerts and the rate $\mathcal{P}(t)$ of total work performed on the particles (consumed power). It is given by $\mathcal{P} (t)\!=\!\sum_n p_n(t)$, where $p_n\!\equiv\!\dot{x}_nF_n\!-\!\dot{x}_nF_n|_{f_n=0}$ and $F_n\!=\!F_n(t)$ is the net athermal force exerted on the particle. For Model~\ref{word:II}, this yields $p_n(t)\!=\![\dot{x}_n(t)\!+\!v u_n (t)]f_n(t)$. These observables attain nonvanishing averages only out of equilibrium.
 In Fig.~\ref{fig:eps}(e), the squared relative uncertainties of all four observables collapse on the universal curve predicted by Eq.~\eqref{eq:eab} in the long-time limit. The above numerical findings motivate a derivation of Eqs.~(\ref{eq:G}-\ref{eq:eab}) from first principles, provided next. 

\section{Theory}
Equations~(\ref{eq:G}-\ref{eq:eab}) follow from the long-time tail of the single-particle conditional probability density (propagator) $P(\mathbf{y},t|\mathbf{y}',0)$. Here, the generalized particle coordinate $\mathbf{y}=(\mathbf{r},\mathbf{u})$ includes both its position $\mathbf{r}$ in a $d$-dimensional space of size $L\rightarrow\infty$ and any internal degrees of freedom $\mathbf{u}$, such as momenta of underdamped particles, active particle orientations or shot noise. The propagator long-time tail, derived below, is~\cite{cn}
\begin{align}
\!\!\!P(\mathbf{y},t|\mathbf{y}',0)  =\frac{\rho_{\text{s}}(\mathbf{y})}{\rho_0 \left(4\pi \Deff t\right)^{d/2}}+\mathcal{O}(t^{-(d/2+1)})\;,\label{eq:prop}
\end{align}
where $\rho_{\text{s}}(\mathbf{y})$
is the steady-state density, which satisfies $\lim_{r \rightarrow\infty}\int d\mathbf{u}\rho_{\rm s}(\mathbf{r},\mathbf{u})\!=\!\rho_0$. The power law in Eq.~\eqref{eq:prop} is a consequence of number conservation~\cite{Pomeau1975,VanBeijeren1982}, which manifests itself for noninteracting particles as probability conservation~\cite{Tyrrell1984,Kubo1991,Mehrer2007}. Indeed, given the initial condition $\rho(\mathbf{y},0)=N \delta(\mathbf{y}-\mathbf{y}')$, the average number density $\rho(\mathbf{y},t)$ is given by $\rho(\mathbf{y},t)=N P(\mathbf{y},t|\mathbf{y}',0)$. The marginal density $\rho(\mathbf{r},t)=\int d\mathbf{u} \rho(\mathbf{r},\mathbf{u},t)$ is governed by Eq.~\eqref{eq:conserv}, which lacks an intrinsic relaxation timescale, rendering $\rho(\br,t)$ a hydrodynamic mode. Moreover, in the absence of interactions, $\rho(\br,t)$ is the only such mode in the system. Equivalently, the position of a particle $\br(t)$ is its only slow degree of freedom.

Equation~\eqref{eq:prop} reveals that the leading order long-time
tail is independent of initial conditions~\footnote{$\rho_{\rm{s}}(\mathbf{y})$ is an infinite invariant density, as it is non-normalizable, i.e., $\int d^d \mathbf{y}\rho_{\rm{s}}(\mathbf{y})=\infty$. For an overview of infinite invariant densities and infinite ergodic theory, see, e.g. Refs.~\cite{Aaronson1997,Aghion2020}}\nocite{Aaronson1997}. It generalizes previous results for $d=1$ BPs~\cite{Miyazawa1999a,Aghion2019,Aghion2020} and RTPs~\cite{Granek2022}. Heuristically, $P(\mathbf{y},t|\mathbf{y}',0)$ is dominated by particle trajectories that avoid long excursions and remain near $\mathbf{y}$. Such particles forget their initial coordinate $\mathbf{y}'$ and relax to a local steady state, i.e $P(\mathbf{y},t|\mathbf{y}',0)\sim\rho_{\rm s}(\mathbf{y})$ up to a normalization factor.  Since the probability density $P(\mathbf{y},t|\mathbf{y}',0)$ expands diffusively at late times, it is supported on a large region of volume $\sim (\Deff t)^{d/2}$. This implies a normalization factor $\sim 1/\rho_0(\Deff t)^{d/2}$.

Equation \eqref{eq:prop} can be applied to obtain the long-time tails of two-time correlation functions of arbitrary stationary and local observables $A(t)=\sum_n a(\mathbf{y}_n(t))$ and $B(t)=\sum_n b(\mathbf{y}_n(t))$, where $\{\mathbf{y}_n\!=\!(\mathbf{r}_n,\mathbf{u}_n)\}_{n=1}^N$ are the generalized coordinates of all particles. Here, local means that $a(\mathbf{y})$ and $b(\mathbf{y})$ decay rapidly for $r=|\mathbf{r}|$ beyond a finite range, with the probe centered at $\mathbf{r}=0$. For $N=\rho_0 L^d$ noninteracting particles, the correlation $\langle A(t)B(0)\rangle _c=\langle A(t)B(0)\rangle{-}\langle A\rangle\langle B\rangle$ is given by $N\langle a\left(\mathbf{y}_1(t)\right)b\left(\mathbf{y}_1(0)\right)\rangle_c$. Since $\langle a\rangle=\langle A\rangle/N$, $\langle A(t)B(0)\rangle _c$ is expressed explicitly as
\begin{align}
\left\langle A(t)B(0)\right\rangle _c	&\!=\!\int\!d\mathbf{y}d\mathbf{y}'\!a(\mathbf{y})P(\mathbf{y},t|\mathbf{y}',0)b(\mathbf{y}')\rho_{\text{s}}(\mathbf{y}'),
\end{align}
with $\mathcal{O}(N^{-1})=\mathcal{O}(L^{-d})$ corrections for finite systems.
Inserting Eq.~\eqref{eq:prop} and $\langle A\rangle\!=\!\int d\mathbf{y}\rho_{\rm s}(\mathbf{y})a(\mathbf{y})$ then leads to
\begin{align}
\!\!\!\left\langle A(t)B(0)\right\rangle _c  \!=\!\frac{\left\langle A\right\rangle \!\left\langle B\right\rangle}{\rho_0 \left(4\pi \Deff t\right)^{d/2}} \!+\!\mathcal{O}(t^{-(d/2+1)}).\label{eq:ratchet}
\end{align}
Setting $B=A$ then yields the first main result, Eq.~\eqref{eq:G}. Furthermore, for $A(t)=\hat{\rho}(\by,t)$ and $B(0)=\hat{\rho}(\by',0)$, where $\hat{\rho}(\by,t)\equiv\sum_n\delta(\by-\by_n(t))$ is the empirical density, Eq.~\eqref{eq:ratchet} provides the density-density correlation function,
\begin{align}
    \!\!\!\left\langle \hat{\rho}(\by,t)\hat{\rho}(\by',0)\right\rangle _c  \!=\!\frac{\rho_{\rm s}(\by)\rho_{\rm s}(\by')}{\rho_0\left(4\pi \Deff t\right)^{d/2}}\!+\!\mathcal{O}(t^{-(d/2+1)})\;,\label{eq:denscorr}
\end{align}
where $\langle\hat{\rho}(\by)\rangle=\rho_{\rm s}(\by)$ is used.

The connection to FCS/DLS can be seen by substituting $A(t)=\int d\by \hat{\rho}(\by,t)a(\by)$ into Eq.~\eqref{eq:Gdef}, which provides
\begin{align}
    G(t)=\frac{\int d\by d\by'a(\by)a(\by')\left\langle \hat{\rho}(\by,t)\hat{\rho}(\by',0)\right\rangle _c}{\left[\int d\by a(\by)\rho_{\rm s}(\by)\right]^2}\;.\label{eq:Gfcs}
\end{align}
Equation~\eqref{eq:Gfcs} is the standard equation in FCS theory, with $\by\!=\!\mathbf{r}$, $\rho_{\rm s}(\br)\!=\!\rho_0$ and $a(\br)$ being the probe's molecule detection efficiency~\cite{Lakowicz2006}. Each FCS variant uses a model for $a(\br)$ in conjunction with the diffusive kernel of noninteracting BPs,
\begin{align}
\langle \hat{\rho}(\br,t)\hat{\rho}(0,0)\rangle_c=\rho_0e^{-r^2/4\Deff t}(4\pi\Deff t)^{-d/2}\;,
\end{align}
leading to various formulae~\cite{Blom2009}, like Eq.~\eqref{eq:FCS} for a Gaussian profile. The alternative proposed here is to expand the kernel at $t\rightarrow\infty$, yielding Eq.~\eqref{eq:G} via Eq.~\eqref{eq:Gfcs}, a result which follows more generally from Eq.~\eqref{eq:denscorr}. In DLS, Eq.~\eqref{eq:Gfcs} provides the contribution of particle number fluctuations to $G(t)$~\cite{Berne2000}. This contribution dominates over $|\mathcal{F}(\mathbf{q},t)|^2\!=\!\beta e^{-2\Deff q^2 t}$ for $t\rightarrow\infty$, again recovering Eq.~\eqref{eq:G}.

The second main result, Eq.~\eqref{eq:eab}, follows from Eq.~\eqref{eq:G} and the identity $\varepsilon^2(t)\!=\!2t^{-2}\!\int_0^t\!ds\!\int_0^{s}\!ds'G(s')$ (time-translation invariance). For $d\leq2$, $\varepsilon^2(t)$ decays slower than $\sim t^{-1}$ due to the nonintegrability of $G(t)\sim t^{-d/2}$ and the finite readout $\langle A\rangle\neq0$. The shift in qualitative behavior for $d\leq2$ results from recurrence -- a non-negligible probability for particles to return to the probe and contribute to $A(t)$ repeatedly~\cite{Erdos1951,Redner2001,Majumdar2024} (see Fig.~\ref{fig:DLSFCS}(a)). The behavior can be interpreted as anomalous diffusion, where strong correlations cause $\langle\bar{A}^2(t)\rangle_c$ to grow nonlinearly with time~\cite{Mandelbrot1968,Dechant2014,Granek2022}.

\subsection{Derivation of Eq.~\eqref{eq:prop}}
The derivation follows the eigenstate expansion approach devised for a one-dimensional BP~\cite{Aghion2019,Aghion2020}. Here, the derivation applies to a generic continuous-time Markov process
satisfying two minimal assumptions: (\textbf{\setword{A1}{word:A1}}) large-scale diffusive behavior to leading order in gradients, and (\textbf{\setword{A2}{word:A2}}) separation of time scales such that the position $\br(t)$ is the only slow degree of freedom. The derivation has two steps: (i) truncate the spectral expansion of the Markov generator at the level of diffusive soft modes, and (ii) invert the expansion to obtain Eq.~\eqref{eq:prop}.

The derivation begins with the general master equation governing the Markov dynamics,
\begin{align}
\frac{d}{dt}\left|P(t)\right\rangle  & =M\left|P(t)\right\rangle ,\label{eq:mast}
\end{align}
where $M$ is a Markov generator and $\langle \mathbf{y}|P(t)\rangle$ specifies the probability density at  $\mathbf{y}$. The solution to Eq.~\eqref{eq:mast} is $|P(t)\rangle   \!=\!U(t)|P(0)\rangle $, where $U(t)=e^{tM}$ is the time-evolution operator. Using the right eigenvectors
$|n\rangle $, left eigenvectors $\langle n|$
and eigenvalues $\epsilon_{n}$, $U(t)$ can be spectrally decomposed as
\begin{align}
U(t) & =\sum_{n=0}^{\infty}e^{-\epsilon_{n}t}\left|n\right\rangle \left\langle n\right|.\label{eq:U}
\end{align}
The ground state $|0\rangle$ corresponds to the steady-state distribution of a single particle $P_{\rm{s}}(\mathbf{y})=\langle \mathbf{y}|0\rangle$, which relates to the steady-state density $\rho_{\rm{s}} (\mathbf{y})$ through $\rho_{\rm{s}} (\mathbf{y})=NP_{\rm{s}}(\mathbf{y})$. Due to the Markov property of $M$, it holds that $\epsilon_0=0$ and ${\rm Re}\,\epsilon_n\geq {\rm Re}\,\epsilon_{n-1}$ for any $n>0$. Therefore, the contributions to Eq.~\eqref{eq:U} due to $|n\rangle\langle n|$, $n\gg1$, become exponentially suppressed in the long-time limit. In this limit, it is thus sufficient to truncate the expansion at the small-${\rm Re}\epsilon_n$ part of the spectrum.

To this end, Assumption~(\ref{word:A1}) provides that, sufficiently far away from the origin, for every solution $|P(t)\rangle$,
\begin{align}
\left\langle \mathbf{r}\left|M\right|P\right\rangle & = D_{ij}\partial_{i}\partial_{j}\left\langle \mathbf{r}|P\right\rangle+\mathcal{O}\!\left[\partial_\mathbf{r}^3\left\langle \mathbf{r}|P\right\rangle\right],\label{eq:Mdiff}
\end{align}
where $\langle\mathbf{r}|\equiv\int d \mathbf{u} \langle\mathbf{y}|$ projects onto the spatial component, $\partial_i \equiv\partial_{r_i}$ and $D_{ij}$ is the effective macroscopic diffusivity tensor. Without loss of generality, the latter is taken to be isotropic, i.e. $D_{ij}=\Deff\delta_{ij}$ where $\delta_{ij}$ is the Kronecker delta. The general anisotropic case can be mapped to the isotropic one by diagonalization and rescaling of the principal axes, yielding $\Deff=(\det D_{ij})^{1/d}$.
Equation~\eqref{eq:Mdiff} implies that, in the limit $L\rightarrow\infty$, the lowest excited states are diffusive modes, whose spatial projections have vanishing spectral gaps~\footnote{Any bound states maintain a finite gap as $L\rightarrow\infty$}. Since the system is inhomogeneous in the vicinity of the probe at $\br=0$, the diffusive modes are given by the standard plane-wave scattering form~\cite{Griffiths2018}
\begin{align}
\left\langle\mathbf{r}|n\right\rangle \sim & \frac{1}{L^d}\left[e^{-i\mathbf{q}\cdot\mathbf{r}}+f(\hat{\mathbf{q}},\hat{\mathbf{r}})\frac{e^{iqr}}{r^{(d-1)/2}}\right]\;,\quad r\rightarrow\infty\;,\label{eq:diffn}
\end{align}
where $\mathbf{q}=\mathbf{q}(n)$ solves $\epsilon_n=\Deff q^2$, $\hat{\mathbf{r}}\equiv\mathbf{r}/r$ and $f(\hat{\mathbf{q}},\hat{\mathbf{r}})$ is the scattering amplitude, which encodes the microscopic details of the system-probe interaction. As demonstrated below, it contributes merely at subleading order in the long-time limit.

To obtain the complete lowest levels $|n\rangle$, Assumption~(\ref{word:A2}) is used. Specifically, for a particle described by the stochastic generalized coordinate $\by(t)=(\br(t),\mathbf{u}(t))$, $\mathbf{u}(t)$ is a fast variable that relaxes at a characteristic rate $\alpha>0$. This allows adiabatic elimination, which provides that, for $t\gg\alpha^{-1}$, any solution $|P(t)\rangle$ is entrained to its spatial projection via the linear mapping~\cite{VanKampen1985a,Gardiner1985},
\begin{align}
\langle\mathbf{y}\left|P(t)\right\rangle&\simeq W\left[\langle\mathbf{r}\left|P(t)\right\rangle\right](\mathbf{y})\;,\label{eq:elim}
\end{align}
where $W$ is an operator determined by $M$~\cite{proj}. For example, when $\mathbf{u}$ and $\mathbf{r}$ are decoupled, $W[g(\mathbf{r})](\mathbf{y})=P_{\rm s}(\mathbf{u})g(\mathbf{r})$, where $P_{\rm s}(\mathbf{u})=\int d\mathbf{r}P_{\rm s}(\mathbf{y})$.
Applying Eq.~\eqref{eq:elim} to the solutions $|P(t)\rangle=e^{-\epsilon_n t}|n\rangle$ provides that the lowest level $|n\rangle$ with a given projection $\langle\mathbf{r}|n\rangle$ is $\langle\mathbf{y}|n\rangle=W\left[\langle\mathbf{r}|n\rangle\right](\mathbf{y})$, such that $\epsilon_n=\Deff q^2 \ll \alpha$. In combination with Eq.~\eqref{eq:diffn}, this allows indexing of the lowest levels by $\mathbf{q}$, i.e., define $|\mathbf{q}\rangle\equiv L^{d}|n\rangle$ and $\langle\mathbf{q}|\equiv \!\langle n|$. The ground state is then $|\mathbf{q}=0\rangle=L^{d}|n=0\rangle$.

In the long-time limit,
only the first terms in Eq.~\eqref{eq:U} corresponding to the diffusive soft modes $|\mathbf{q}\rangle$ contribute, providing
\begin{align}
\!\!\!\!U(t)\!\sim\!\frac{1}{L^d}\sum_{\mathbf{q}}e^{-\epsilon(\mathbf{q})t}\left|\mathbf{q}\right\rangle \!\left\langle \mathbf{q}\right|\!\sim\!\int\frac{d^{d}\mathbf{q}}{(2\pi)^{d}}e^{-\epsilon(\mathbf{q})t}\left|\mathbf{q}\right\rangle \!\left\langle \mathbf{q}\right|,\label{eq:sp}
\end{align}
where $\epsilon(\mathbf{q})=\Deff q^2$ and the convergence to an integral is obtained for $L\rightarrow\infty$.
The last step is a saddle-point
approximation of Eq.~\eqref{eq:sp} in the limit $t\rightarrow\infty$, which amounts to expanding $|\mathbf{q}\rangle$ as $|\mathbf{q}\rangle=|0\rangle+\mathcal{O}(q)$ and evaluating the Gaussian integral. The result is
\begin{align}
U(t) & =\left|0\right\rangle \!\left\langle 0\right|\left(4\pi \Deff t\right)^{-d/2}+\mathcal{O}(t^{-(d/2+1)})\;.\label{eq:propspec}
\end{align}
In the basis $\{|\mathbf{y}\rangle\}$, it holds that $P(\mathbf{y},t|\mathbf{y}',0)\!\equiv\!\langle \mathbf{y}|U(t)|\mathbf{y}'\rangle$, $\langle 0|\mathbf{y}'\rangle\!=\!1$ and $\langle \mathbf{y}|0\rangle\!=\!L^d P_{\text{s}}(\mathbf{y})$. The factor of $L^d$ stems from the redefinition of $|0\rangle$ in Eq.~\eqref{eq:sp}. Using $\rho_{\rm{s}} (\mathbf{y})\!=\!\rho_0 L^{d} P_{\rm{s}}(\mathbf{y})$ then leads to Eq.~\eqref{eq:prop}.

Importantly, the $\mathcal{O}(q)$ corrections to the Gaussian integral in Eq.~\eqref{eq:sp} lead to $\mathcal{O}(t^{-(d/2+1)})$ corrections within the saddle-point expansion in Eq.~\eqref{eq:propspec} and hence in Eq.~\eqref{eq:prop}. The derivation of the prefactor of the leading $\sim t^{-(d/2+1)}$ correction is provided in the Appendix, where it is shown to be nonuniversal.

\section{Discussion and conclusions}
In this article, the universal Eqs.~(\ref{eq:G}-\ref{eq:eab}) are derived for local probe measurements in dilute diffusive systems.
Together, these equations permit direct local measurement of $\Deff$ in microscopic regions \textemdash{} where Fick's law may fail \textemdash{} and apply broadly across observables and nonequilibrium steady states within the dilute diffusive regime. They can promote existing indirect methods, such as DLS and FCS, which face challenges in short-time measurement of active systems~\cite{Boon1974,Wilson2011,Gunther2018}, to direct methods. The only external input required is the macroscopic state variable $\rho_0$.
Conversely, if $\Deff$ is known, the formulae allow direct local measurement of $\rho_0$. Put differently, Eq.~\eqref{eq:G} implies an equation-of-state-like relation,
\begin{align}
    G_\infty(\rho_0,\Deff)=\frac{1}{\rho_0(4\pi\Deff)^{d/2}}\;,
\end{align}
where $G_\infty\equiv\lim_{t\rightarrow\infty}t^{d/2}G(t)$.
These predictions may motivate experimental strategies to overcome current barriers to long-term probing~\cite{Berne2000,Lakowicz2006,Gunther2018} and the development of novel local probes that measure nonconventional observables. The experimental realization of these possibilities remains open.
Several implications of Eq.~\eqref{eq:eab}, as well as extensions and relations to prior results, are discussed below.

\subsection{Consequences for thermodynamic bounds}
 For a broad class of systems, a measurement $A(t)$ that is odd under time reversal, such as current or power, satisfies the thermodynamic uncertainty relation~\cite{Seifert2019},
 \begin{align}
     \varepsilon^2\geq\frac{2}{\Sigma t}\;,\label{eq:TUR}
 \end{align}
where $\Sigma$ is the average entropy production rate. Conversely, if $A(t)$ is even under time reversal, it satisfies the kinetic uncertainty relation~\cite{DiTerlizzi2019},
\begin{align}
     \varepsilon^2\geq\frac{1}{\mathcal{A} t}\;,\label{eq:KUR}
 \end{align}
 where $\mathcal{A}$ is the average dynamical activity.  So-called ``hyperaccurate'' observables can approach the bounds in Eqs.~(\ref{eq:TUR}-\ref{eq:KUR})~\cite{Busiello2019,VanVu2020,Shiraishi2021,Busiello2022,Dieball2023,Timpanaro2023} and thereby assist in inferring $\Sigma$ and $\mathcal{A}$ from empirical data~\cite{Seifert2019,VanVu2020,Dieball2023}. However, Eq.~\eqref{eq:eab} demonstrates a decay slower than $1/t$, indicating that for $d\leq2$, microscopic local measurements in macroscopic systems cannot saturate the uncertainty relations. The asymptotic distance to the bound grows as a power law, independent of the local observable. Further extensions and relations are collected next.

\subsection{Extensions and relation to prior results}
The extension of Eqs.~(\ref{eq:G}-\ref{eq:eab}) to dense systems, where interactions cannot be neglected, remains an open direction. Similar results may hold with $\Deff$ renormalized by the interactions. Generalization to other transport coefficients, such as viscosity or thermal conductivity, also remains to be explored.

The presented theory offers a bridge between the transient power-law relaxations obtained previously and long-time correlations in stationary systems~\cite{Agmon1988,Miyazawa1999a,Agmon2011,Simkovitch2016,Sivan2018,Aghion2019,Aghion2020,Defaveri2023}. The extension of these previous results to stationary correlations remains open for future investigation.

Equation~\eqref{eq:eab} contrasts with earlier instances of universal fluctuations. The Darling–Kac theorem establishes a universal asymptotic distribution for nonstationary, nonnegative local observables, with a dependence solely on  $d$~\cite{Darling1957,Aaronson1997,Aghion2019}. By contrast, Eq.~\eqref{eq:eab} reveals a form of steady-state universality that holds for observables of arbitrary sign and depends explicitly on the macroscopic parameters $\rho_0$ and $\Deff$. The possible extension
of this universality to rare fluctuations, in analogy with the Darling-Kac theorem, remains an open question. Interestingly, relative fluctuations that are independent of microscopic details also emerge in biased random walks with broadly distributed quenched disorder~\cite{Bouchaud1990a,Bouchaud1990b}.

Equations~(\ref{eq:G}-\ref{eq:eab}) admit several direct extensions. First, Eq.~\eqref{eq:ratchet} applies to both autocorrelations and cross-correlations. Accordingly, the forms in Eqs.~(\ref{eq:G}-\ref{eq:eab}) extend directly to $G_{AB}(t)\!\equiv\!\langle A(t)B(0)\rangle/\langle A\rangle\langle B\rangle$ and $\varepsilon_{AB}^2(t)\!\equiv\!\langle \bar{A} (t) \bar{B} (t)\rangle_{\rm c}/\langle \bar{A}\rangle\langle \bar{B}\rangle$, respectively. Second, the present results apply to fixed probes. Their extension to dynamic probes (e.g. tracer particles~\cite{Seyforth2022,Venturelli2023}) is expected to require adiabatic expansions~\cite{VanKampen1986,DAlessioJSTAT2016,Maes2020,Granek2022,Solon2022}.

The equations are also expected to hold in inhomogeneous systems, such as those with disordered or periodic potentials, where $\Deff$ is renormalized by the microscopic structure~\cite{Lifson1962,Derrida1983,Vergassola1997,Dean2007}. This is consistent with the observed transient power laws~\cite{Sivan2018,Defaveri2023}. Furthermore, inhomogeneous boundary conditions or bulk inhomogeneity can provide a macroscopically inhomogeneous $\Deff(\br)$, which may be detected by scanning probe positions. These extensions break when inhomogeneity generates ballistic motion at a finite average velocity $\mathbf{U}$, thus violating the effective diffusion Assumption~(\ref{word:A1}). In this case, the algebraic decay is lost at a cut-off timescale $\sim\Deff/U^2$, leading to exponential relaxation at a rate $U^2/\Deff$~\cite{Luck2001}.

Finally, the relationship between the present findings and the well-known long-time tails in heat and momentum transport~\cite{Pomeau1975,VanBeijeren1982,Dhar2008a} remains to be explored. Remarkably, assuming Eq.~\eqref{eq:G} with $\Deff$ being the thermal diffusivity and $N$ the effective number of degrees of freedom reproduces the long-time energy ACF in an idealized model of thermal transport~\cite{Ernst2005,Ripoll2005}, hinting toward deeper connections.

\begin{acknowledgments}
I thank my PhD advisor, Yariv Kafri, for his support and many educating and insightful discussions. I thank Julien Tailleur, Shlomi Reuveni, Eli Barkai, Thomas A. Witten, Felix H{\"o}fling, Rony Granek, Yael Avni and Ran Yaacoby for helpful discussions. I thank the anonymous referees for helpful comments. I acknowledge support from the Leinweber Institute for Theoretical Physics and the Center for Living Systems at The University of Chicago. I acknowledge support from a
MRSEC-funded Kadanoff–Rice fellowship and The University of Chicago Materials Research Science and Engineering Center, which is funded by NSF (DMR-2011854). I acknowledge support from ISF (2038/21), NSF/BSF (2022605) and the Adams Fellowship Program of the Israeli Academy of Sciences and Humanities. 
\end{acknowledgments}

\appendix*
\section{Sub-leading corrections}\label{app:subleading}
Here, the derivation of Eqs.~\eqref{eq:G} and~\eqref{eq:prop} of the main text is extended to include the leading-order correction in the long-time limit. Importantly, it is shown that this correction is non-universal and depends on microscopic details. For completeness and clarity, the derivation is reproduced in its entirety and extended where appropriate. The derivation follows the eigenstate expansion approach devised for a one-dimensional BP~\cite{Aghion2019,Aghion2020}. Here, the derivation applies to a generic continuous-time Markov process satisfying two minimal assumptions: (\ref{word:A1}) large-scale diffusive behavior to leading order in gradients, and (\ref{word:A2}) separation of time scales such that the position $\br(t)$ is the only slow coordinate.

The derivation begins with the general master equation governing the Markov dynamics,
\begin{align}
    \frac{d}{dt}|P(t)\rangle &= M |P(t)\rangle ,\label{eq:app_mast}
\end{align}
where $M$ is a Markov generator and $\langle\by|P(t)\rangle$ specifies the probability density at $\by$. The solution to Eq.~\eqref{eq:app_mast} is $|P(t)\rangle=U(t)|P(0)\rangle$, where $U(t)=e^{tM}$ is the time-evolution operator. Using the right eigenvectors $|n\rangle$, left eigenvectors $\langle n|$ and eigenvalues $\epsilon_n$, $U(t)$ can be spectrally decomposed as
\begin{align}
U(t) & =\sum_{n=0}^{\infty}e^{-\epsilon_{n}t}\left|n\right\rangle \left\langle n\right|.\label{eq:app_U}
\end{align}
The ground state $|0\rangle$ corresponds to the steady-state distribution of a single particle $P_{\rm{s}}(\by)=\langle \by|0\rangle$, which relates to the steady-state density $\rho_{\rm{s}}(\by)$ through $\rho_{\rm{s}}(\by)=NP_{\rm{s}}(\by)$. Due to the Markov property of $M$, it holds that $\epsilon_0=0$ and ${\rm Re}\,\epsilon_n\geq {\rm Re}\,\epsilon_{n-1}$ for any $n>0$. Therefore, the contributions to Eq.~\eqref{eq:app_U} due to $|n\rangle\langle n|$, $n\gg1$, become exponentially suppressed in the long-time limit. In this limit, it is thus sufficient to truncate the expansion at the small-${\rm Re}\epsilon_n$ part of the spectrum.

To this end, Assumption~(\ref{word:A1}) is invoked. Sufficiently far from the origin and over sufficiently large length scales, the projection of the dynamics onto the Euclidean space $\{\br\}=\mathbb{R}^d$ is local. Therefore, the operator $\mathcal{M}(\br)$ defined by
\begin{align}
    \partial_t\langle\br|P\rangle = \langle \br |M| P\rangle \equiv \mathcal{M}\langle\br|P\rangle ,\label{eq:app_Mdef}
\end{align}
where $\langle\br|\equiv \int d\mathbf{u}\langle\by|$ projects onto the spatial component, admits the general gradient expansion
\begin{align}
\mathcal{M}(\partial_\br) &= M_0 - V_a\partial_a + D_{ab}\partial_a\partial_b + K_{abc}\partial_a\partial_b\partial_c \nonumber\\
&\quad + C_{abcd}\partial_a\partial_b\partial_c\partial_d + \mathcal{O}(\partial_\br^5),\label{eq:app_grad}
\end{align}
where $\partial_a\equiv \partial_{r_a}$. Conservation of probability imposes $M_0=0$. Furthermore, Assumption~(\ref{word:A1}) excludes the case of ballistic transport, thus enforcing $V_a=0$ and reproducing Eq.~\eqref{eq:Mdiff} as the leading order expansion.

The effective macroscopic diffusivity tensor $D_{ab}$ is symmetric. Thus, the quadratic form in Eq.~\eqref{eq:app_grad} can be diagonalized so that $D_{ab}\rightarrow D_a\delta_{ab}$ by a linear change of variables $r_a\rightarrow O_{ab}r_b$, where $O_{ab}$ is an orthogonal matrix. Furthermore, by introducing $D_a\equiv \Deff I_a$ such that $\prod_{a=1}^d I_a=1$ and $\Deff=(\det D_{ab})^{1/d}$, the rescaling $r_a\rightarrow r_a/I_a^{1/2}$ yields $D_{ab}\rightarrow \Deff\delta_{ab}$. Equation~\eqref{eq:app_grad} then becomes
\begin{align}
 \mathcal{M}(\partial_\br) &= \Deff \partial_\br^2+\mathcal{L}(\partial_\br),\label{eq:app_Miso}\\
 \mathcal{L}(\partial_\br) &\equiv K_{abc}\partial_a\partial_b\partial_c + C_{abcd}\partial_a\partial_b\partial_c\partial_d + \mathcal{O}(\partial_\br^5).\label{eq:app_L}
\end{align}
For the eigenbasis $\{|n\rangle\}$, Eq.~\eqref{eq:app_Mdef} provides the eigenvalue problem
\begin{align}
    -\epsilon_n\langle\br|n\rangle = \mathcal{M}(\partial_\br)\langle\br|n\rangle .\label{eq:app_eig}
\end{align}
In the limit $L\rightarrow\infty$, the lowest excited states are thus diffusive modes, whose spatial projections have vanishing spectral gaps. Any bound states maintain a finite gap as $L\rightarrow\infty$. Since the system is inhomogeneous in the vicinity of the probe at $\br=0$, the solutions $\langle\br|n\rangle$ admit the general plane-wave scattering decomposition
\begin{align}
    \langle\br|n\rangle &= \frac{1}{L^d}\left[e^{-i\mathbf{q}\cdot\br}+\psi(\mathbf{q},\br)\right],\label{eq:app_psi}
\end{align}
where the scattering term $\psi(\mathbf{q},\br)$ satisfies $\lim_{r\rightarrow\infty}\psi(\mathbf{q},\br)=0$. Substituting Eq.~\eqref{eq:app_psi} in Eq.~\eqref{eq:app_eig} and taking $r\rightarrow\infty$ then yields that $\mathbf{q}=\mathbf{q}(n)$ solves the dispersion relation
\begin{align}
    \epsilon_n = -\mathcal{M}(-i\mathbf{q}) = \Deff q^2 - \mathcal{L}(-i\mathbf{q}).\label{eq:app_dispersion}
\end{align}
The functional form of $\psi(\mathbf{q},\br)$ depends on the microscopic details of the system-probe interaction. At leading order in $r\rightarrow\infty$, where $\mathcal{L}(\partial_\br)$ can be neglected in Eq.~\eqref{eq:app_eig}, it is given by the standard scattering expansion
\begin{align}
    \psi(\mathbf{q},\br) &\sim f(\hat{\mathbf{q}},\hat{\br})\frac{e^{iqr}}{r^{(d-1)/2}},\qquad r\rightarrow\infty .\label{eq:app_scatter}
\end{align}
where the scattering amplitude $f(\hat{\mathbf{q}},\hat{\br})$ encodes the microscopic details. Note that a similar decomposition exists for $\langle n|\br\rangle$,
\begin{align}
    \langle n|\br\rangle &= \frac{1}{L^d}\left[e^{i\mathbf{q}\cdot\br}+\phi(\mathbf{q},\br)\right],\label{eq:app_phi}
\end{align}
where $\lim_{r\rightarrow\infty}\phi(\mathbf{q},\br)=0$.

To obtain the complete low part of the spectrum, Assumption~(\ref{word:A2}) is used. Specifically, for a particle described by the stochastic generalized coordinate $\by(t)=(\br(t),\mathbf{u}(t))$, $\mathbf{u}(t)$ is a fast variable that relaxes at a characteristic rate $\alpha>0$. This allows for adiabatic elimination of $\mathbf{u}(t)$ as follows.

First, $|P(t)\rangle$ can be decomposed as
\begin{align}
    |P(t)\rangle = \Pi|P(t)\rangle + (1-\Pi)|P(t)\rangle ,\label{eq:app_proj}
\end{align}
where $\Pi=\int d\br P_{\rm{s}}(\mathbf{u}|\br)|\br\rangle\langle\br|$ is a projection operator and $P_{\rm{s}}(\mathbf{u}|\br)$ is the steady-state distribution of $\mathbf{u}(t)$ conditioned on the particle being fixed at $\br(t)=\br$. It can be shown that, for $t\gg\alpha^{-1}$, $(1-\Pi)|P\rangle\simeq R\Pi|P\rangle$ where $R(\alpha)$ is a time-independent linear operator~\cite{VanKampen1985a,Gardiner1985}. It thus follows that, for $t\gg\alpha^{-1}$,
\begin{align}
    |P(t)\rangle \simeq (1+R)\Pi|P(t)\rangle ,\label{eq:app_elim}
\end{align}
reproducing Eq.~\eqref{eq:elim} with $W[\langle\br|P(t)\rangle]=(1+R)\Pi|P(t)\rangle$.

Applying Eq.~\eqref{eq:app_elim} to the solutions $|P(t)\rangle=e^{-\epsilon_n t}|n\rangle$ provides
\begin{align}
    e^{-\epsilon_n t}|n\rangle \simeq e^{-\epsilon_n t}(1+R)\Pi|n\rangle .\label{eq:app_elimn}
\end{align}
Since Eq.~\eqref{eq:app_elimn} holds for $t\gg\alpha^{-1}$, both sides of the equality are nonvanishing only when $\epsilon_n\ll\alpha$, i.e. $|n\rangle$ is a soft mode whose eigenvalue lies below the spectral gap $\alpha$ of the internal dynamics. Thus, $|n\rangle=(1+R)\Pi|n\rangle=W[\langle\br|n\rangle]$ such that Eqs.~\eqref{eq:app_psi}-\eqref{eq:app_dispersion} hold. This allows to index the soft modes by $\mathbf{q}$, i.e., define $|\mathbf{q}\rangle\equiv L^d|n\rangle$ and $\langle\mathbf{q}|\equiv \langle n|$. The ground state is then $|\mathbf{q}=0\rangle=L^d|n=0\rangle$.

In the long-time limit, only the first terms in Eq.~\eqref{eq:app_U} corresponding to the diffusive soft modes $|\mathbf{q}\rangle$ contribute, providing
\begin{align}
U(t)&\sim\frac{1}{L^d}\sum_{\mathbf{q}}e^{-\epsilon(\mathbf{q})t}|\mathbf{q}\rangle\langle\mathbf{q}|\sim\int\frac{d^d\mathbf{q}}{(2\pi)^d}e^{-\epsilon(\mathbf{q})t}|\mathbf{q}\rangle\langle\mathbf{q}|,\label{eq:app_sp}
\end{align}
where $\epsilon(\mathbf{q})=\Deff q^2-\mathcal{L}(-i\mathbf{q})$ and the convergence to an integral is obtained for $L\rightarrow\infty$. The last step is a saddle-point approximation of Eq.~\eqref{eq:app_sp} in the limit $t\rightarrow\infty$, which amounts to expanding $Q(\mathbf{q})\equiv|\mathbf{q}\rangle\langle\mathbf{q}|$ and $e^{\mathcal{L}(-i\mathbf{q})t}$ around $\mathbf{q}=0$ and evaluating the Gaussian integral. The result is the asymptotic power-series expansion
\begin{align}
U(t)&=(4\pi\Deff t)^{-d/2}\left[|0\rangle\langle0|+\frac{U_1(0)}{t}+\mathcal{O}(t^{-2})\right],\label{eq:app_U1exp}
\end{align}
where
\begin{align}
U_1(\mathbf{q})&=\frac{\partial_{\mathbf{q}}^2 Q}{4\Deff}+\frac{3iK_{abb}\partial_{q_a}Q}{4\Deff^2}+\frac{3C_{aabb}Q}{4\Deff^2}\nonumber\\
&\quad -\frac{3(2K_{abc}K_{abc}+3K_{aac}K_{bbc})Q}{16\Deff^3}.\label{eq:app_U1}
\end{align}
In the basis $\{|\by\rangle\}$, it holds that $P(\by,t|\by',0)\equiv\langle\by|U(t)|\by'\rangle$, $\langle0|\by'\rangle=1$ and $\langle\by|0\rangle=L^dP_{\rm{s}}(\by)$. The factor of $L^d$ stems from the redefinition of $|0\rangle$ in Eq.~\eqref{eq:app_sp}. Using $\rho_{\rm{s}}(\by)=\rho_0L^dP_{\rm{s}}(\by)$ then leads to
\begin{align}
P(\by,t|\by',0)&=\frac{1}{\rho_0(4\pi\Deff t)^{d/2}}\nonumber\\&\quad\cdot\left[\rho_{\rm{s}}(\by)+\frac{\rho_1(\by,\by')}{t}+\mathcal{O}(t^{-2})\right],\label{eq:app_prop1}
\end{align}
where $\rho_1(\by,\by')\equiv \rho_0\langle\by|U_1(0)|\by'\rangle$ is given by
\begin{align}
\!\!\!\!\rho_1(\by,\by')&=\frac{Q^{(2)}(\by,\by')}{4\Deff}+\frac{3iK_{abb}Q_a^{(1)}(\by,\by')}{4\Deff^2}\nonumber\\
&\!\!+\left(\frac{3C_{aabb}}{4\Deff^2}-\frac{6K_{abc}K_{abc}+9K_{aac}K_{bbc}}{16\Deff^3}\right)\!\rho_{\rm{s}}(\by),\label{eq:app_rho1}\\
\mathbf{Q}^{(1)}(\by,\by')&\equiv \rho_0\langle\by|\partial_{\mathbf{q}}Q|\by'\rangle|_{\mathbf{q}=0},\label{eq:app_Q1}\\
Q^{(2)}(\by,\by')&\equiv \rho_0\langle\by|\partial_{\mathbf{q}}^2Q|\by'\rangle|_{\mathbf{q}=0}.\label{eq:app_Q2}
\end{align}
Thus, $\rho_1(\by,\by')$ provides the leading-order correction to Eq.~\eqref{eq:prop}. Inserting Eq.~\eqref{eq:app_prop1} into Eq.~\eqref{eq:ratchet} then leads to
\begin{align}
G(t)&=\frac{1}{\rho_0(4\pi\Deff t)^{d/2}}\left[1+\frac{G_1}{t}+\mathcal{O}(t^{-2})\right],\label{eq:app_G1form}
\end{align}
where
\begin{align}
G_1&=\frac{E^{(2)}}{4\Deff}+\frac{3iK_{abb}E_a^{(1)}}{4\Deff^2}+\frac{3C_{aabb}}{4\Deff^2}\nonumber\\&-\frac{6K_{abc}K_{abc}+9K_{aac}K_{bbc}}{16\Deff^3},\label{eq:app_G1}\\
\mathbf{E}^{(1)}&\equiv \rho_0\frac{\langle A|\partial_{\mathbf{q}}Q|B\rangle}{\langle A\rangle\langle B\rangle}\nonumber\\
&=\frac{\int d\by d\by' A(\by)\mathbf{Q}^{(1)}(\by,\by')B(\by')\rho_{\rm{s}}(\by')}{\int d\by\rho_{\rm{s}}(\by)A(\by)\int d\by'\rho_{\rm{s}}(\by')B(\by')},\label{eq:app_E1}\\
E^{(2)}&\equiv \rho_0\frac{\langle A|\partial_{\mathbf{q}}^2Q|B\rangle}{\langle A\rangle\langle B\rangle}\nonumber\\
&=\frac{\int d\by d\by' A(\by)Q^{(2)}(\by,\by')B(\by')\rho_{\rm{s}}(\by')}{\int d\by\rho_{\rm{s}}(\by)A(\by)\int d\by'\rho_{\rm{s}}(\by')B(\by')}.\label{eq:app_E2}
\end{align}
Equation~\eqref{eq:app_G1form} provides the leading-order $\mathcal{O}(t^{-(d/2+1)})$ correction to Eq.~\eqref{eq:G}. As evident in Eqs.~\eqref{eq:app_G1}-\eqref{eq:app_E2}, the correction coefficient $G_1$ is nonuniversal. Indeed, since $|\mathbf{q}\rangle\simeq W[\langle\br|\mathbf{q}\rangle]$ with $\langle\br|\mathbf{q}\rangle$ being given by Eq.~\eqref{eq:app_psi}, $G_1$ depends on $\partial_{\mathbf{q}}\psi(\mathbf{q},\br)|_{\mathbf{q}=0}$ and $\partial_{\mathbf{q}}^2\psi(\mathbf{q},\br)|_{\mathbf{q}=0}$. Similarly, since $\langle\mathbf{q}|\simeq W^\dagger[\langle\mathbf{q}|\br\rangle]$, where $W^\dagger$ denotes the dual map under the biorthogonal pairing, $G_1$ also depends on $\partial_{\mathbf{q}}\phi(\mathbf{q},\br)|_{\mathbf{q}=0}$ and $\partial_{\mathbf{q}}^2\phi(\mathbf{q},\br)|_{\mathbf{q}=0}$. In summary, while the leading order long-time tails in Eqs.~(\ref{eq:G}-\ref{eq:eab}) are universal, the subleading corrections are not: they depend on microscopic details of the particle dynamics, the system-probe interaction and the measured observable.

\bibliographystyle{apsrev4-2}
\bibliography{references}

\begin{thebibliography}{112}
\bibitem [{Log()}]{LogAvg}%
  \BibitemOpen
  \href@noop {} {}\bibinfo {note} {A logarithmic moving average of windows
  $[t/1.01,1.01t]$ was further applied to BP DLS data.}\BibitemShut {Stop}%
\bibitem [{pen()}]{pent}%
  \BibitemOpen
  \href@noop {} {}\bibinfo {note} {The wall penetrability does not affect the
  equations of state and the validity of the results.}\BibitemShut {Stop}%
\bibitem [{cn()}]{cn}%
  \BibitemOpen
  \href@noop {} {}\bibinfo {note} {The expansion is taken with
  $t\rightarrow\infty$ and fixed $\mathbf{y}$ and $\mathbf{y}'$, i.e. for
  $r,r'\ll(\Deff t)^{1/2}$. For finite-yet-large $L$, the long-time tails are
  valid for $t\ll L^2 /\Deff$, after which the decay becomes exponential. The
  slow decay then materializes as a dependence of $\mathcal{D}$ on
  $L$.}\BibitemShut {Stop}%
\bibitem [{pro()}]{proj}%
  \BibitemOpen
  \href@noop {} {}\bibinfo {note} {Eq.~\eqref{eq:elim} is equivalent to
  $|P(t)\rangle\simeq W\Pi|P(t)\rangle$, where $\Pi=\int
  d\mathbf{r}|\mathbf{r}\rangle\langle\mathbf{r}|$ is a projection
  operator.}\BibitemShut {Stop}%
\end{thebibliography}

\end{document}